\documentclass[graybox]{svmult}

\usepackage{mathptmx}       % selects Times Roman as basic font
\usepackage{helvet}         % selects Helvetica as sans-serif font
\usepackage{courier}        % selects Courier as typewriter font
\usepackage{type1cm}        % activate if the above 3 fonts are
\usepackage{makeidx}         % allows index generation
\usepackage{graphicx}        % standard LaTeX graphics tool
\usepackage{multicol}        % used for the two-column index
\usepackage[bottom]{footmisc}% places footnotes at page bottom

\makeindex             % used for the subject index
\begin{document}

\title*{Multi-Periodic Oscillations in Cepheids and RR~Lyrae-Type Stars}
\titlerunning{Multi-Periodic Oscillations in Cepheids and RR Lyrae Stars}
\author{P. Moskalik}
\institute{P. Moskalik  \at Copernicus Astronomical Center, Warszawa, Poland,
                            \email{pam@camk.edu.pl}}
\maketitle
% Please use both starred abstract and non-starred abstract.

\abstract*{Classical Cepheids and RR Lyrae-type stars are usually
considered to be textbook examples of purely radial, strictly
periodic pulsators. Not all the variables, however, conform to this
simple picture. In this review I discuss different forms of
multi-periodicity observed in Cepheids and RR~Lyrae stars, including
Blazhko effect and various types of radial and nonradial multi-mode
oscillations.}

\abstract{Classical Cepheids and RR Lyrae-type stars are usually
considered to be textbook examples of purely radial, strictly
periodic pulsators. Not all the variables, however, conform to this
simple picture. In this review I discuss different forms of
multi-periodicity observed in Cepheids and RR~Lyrae stars, including
Blazhko effect and various types of radial and nonradial multi-mode
oscillations.}

\section{Blazhko effect}
\label{sec:2}

Blazhko effect is a slow, nearly periodic modulation of the
pulsation amplitude and phase. It is observed in many RR~Lyrae
stars, both of RRab (fundamental mode) and of RRc type (overtone
pulsators). The modulation period can range from a few days to over
2500 days \cite{BulgeRR}. The phenomenon was discovered over a
century ago (\cite{Blazhko, Shapley}) and was the first identified
departure from the "strictly periodic" paradigm.

Blazhko effect has been detected in various stellar systems,
including Magellanic Clouds \cite{LMCRR,SMCRR}, galactic bulge
\cite{BulgeRR} and several globular clusters (e.g \cite{OmCenStat}).
By no mean it is a rare phenomenon. Modulation occurs in $\sim\!
50$\% of the field RRab stars (\cite{KBS,KepBenko}). The incidence
rates reported in other RR~Lyrae populations are up to 4 times
lower, perhaps because of lower quality of the available data. In
most stellar systems the incidence rate of Blazhko modulation is
higher in RRab stars than in RRc stars. The opposite is true only in
the globular cluster Omega Centauri, where unusually high fraction
(38\%) of modulated RRc stars has been found \cite{OmCenStat}.

Blazhko effect is usually associated with the RR~Lyrae variables,
but it can also occur (albeit rarely) in classical Cepheids. A
modulation with a period of $\sim\! 1200$\thinspace days is observed
in a single-mode overtone Cepheid V473~Lyr \cite{V473Lyr}. The
amplitude and phase modulation has also been recently discovered in
a number of LMC double-mode Cepheids \cite{MK07}. These latter
variables will be discussed in Sec.~\ref{sec:3.2}.

After 100 years since its discovery, the Blazhko effect remains an
unsolved mystery. Several models have been proposed to explain the
phenomenon (e.g. \cite{MORP,NRRP,Stothers}), but all of them have
failed (see excellent review by Kov\'acs \cite{BlazhRev}). Recent
discovery of the period doubling in the Blazhko RR~Lyrae stars
\cite{PDobs,PDrev} offers a new hope of understanding the Blazhko
enigma. The period doubling is a resonant phenomenon
\cite{MosBuch90}. In the RRab models it has been traced to the 9:2
resonance between the fundamental mode and the $9^{\rm th}$ radial
overtone \cite{PDmodel}. In principle, such a coupling is capable of
producing not only stationary, but also modulated solutions
\cite{PDAE}. This makes the 9:2 resonance interaction a very
promising idea, that might explain the Blazhko phenomenon. But we
are not there yet. So far, nonlinear RR~Lyrae models have been able
to reproduce the period doubling, but without the modulation
\cite{PDmodel}. It remains to be seen if {\it modulated} period
doubling solutions can also be found.

\section{Multi-mode radial pulsators}
\label{sec:3}

\subsection{Double-mode RR Lyrae-type star (RRd stars)}
\label{sec:3.1}

AQ~Leo, the first double-mode RR~Lyrae-type variable, was identified
only in 1977 \cite{AQLeoJW}. Since then, double-mode RR~Lyrae
pulsators have been identified not only in the galactic field, but
also in several globular clusters (e.g. \cite{M15RR,IC4499RR,M68RR})
and in many dwarf spheroidal galaxies (e.g.
\cite{TucRR,ForRR,SgrRR,DraRR,SclRR}). 3 such stars are also known
in M31 \cite{M31RR}. By far the largest populations of RRd stars has
been found in the Magellanic Clouds: 986 variables in the LMC
\cite{LMCRR} and 258 variables in the SMC \cite{SMCRR}. For
comparison, in the Galaxy we know only about 90 RRd stars in the
field \cite{GalRR} and 80 in the bulge \cite{BulgeRR}. Typically,
double-mode pulsators constitute several per cent of the RR~Lyrae
star population, but in some stellar systems this fraction can be as
high as $20-30$\% \cite{ForRR,M68RR} or as low as $0.5-1$\%
\cite{SzczFab,BulgeRR}.

In vast majority of the RRd stars the period ratio of the two
excited modes, $P_1/P_0$, is in a narrow range of $0.742-0.748$.
This identifies the two modes as the first radial overtone (1O) and
the radial fundamental mode (F). In most cases, the first overtone
has larger amplitude than the fundamental (see e.g. \cite{LMCRR}).
When plotted on the $P_1/P_0$ vs. $P_0$ plane (the so-called
Petersen diagram), the RRd stars form a well defined sequence, with
the period ratio being systematically higher at longer period (e.g.
\cite{LMCRR}, their Fig.\thinspace 4). The exact location of the
star in this plot is determined mainly by its metallicity
\cite{Popiel,SclRR}. In most cases, the observed period ratios
correspond to [Fe/H] between $-1.3$ and $-2.0$. Only in the unique
case of the galactic bulge, the RRd sequence extends down to
$P_0=0.35$\thinspace day and $P_1/P_0=0.726$, which implies
metallicities as high as [Fe/H]$=-0.35$ \cite{BulgeRR}.

\subsection{Double-mode Cepheids}
\label{sec:3.2}

First double-mode Cepheids have been discovered more than half a
century ago \cite{UTrA,TUCas}. These type of pulsators come in two
basic flavours: they either pulsate in the fundamental mode and the
first overtone (F/1O type, $P_1/P_0=0.695-0.745$) or in the first
two radial overtones (1O/2O type, $P_2/P_1=0.79-0.81$). Recently, a
third type of double-mode Cepheids has been discovered
\cite{LMCTripl}, with the first and the third overtones
simultaneously excited (1O/3O type, $P_3/P_1=0.677$). So far, only
two such stars have been identified.

Almost 700 double-mode Cepheids are currently known. The largest
populations have been found in the LMC \cite{LMCCep,EROSCep} (90
F/1O + 256 1O/2O + 2 1O/3O) and in the SMC \cite{SMCCep} (59 F/1O +
215 1O/2O). For comparison, in the Galaxy we know only 24 F/1O
Cepheids and 16 1O/2O Cepheids. 5 F/1O Cepheids are known in M33
\cite{M33Cep}.

The values of $P_2/P_1$ are almost the same for all the 1O/2O
Cepheids in all stellar systems. Behaviour of $P_1/P_0$ is
different. This period ratio becomes systematically lower as the
pulsation period increases. It also differs between stellar systems,
being highest in the SMC, intermediate in the LMC and lowest in the
Galaxy (e.g. \cite{MK07}, their Fig.\thinspace 1). The latter
property is caused by a strong metallicity dependence of $P_1/P_0$.

\subparagraph{Blazhko effect in 1O/2O double-mode Cepheids}
\label{sec:3.2.1}

Analysis of MACHO and OGLE-II photometry of LMC variables has led to
discovery of a Blazhko effect in 1O/2O Cepheids \cite{MK07}. At
least 20\% of these double-mode pulsators display periodic
modulations. Amplitudes and phases of both modes vary, with a common
period, $P_B$, which is always longer than 700\thinspace day. The
longest possible modulation period is not known, currently it is
limited only by the length of the data. Modulation is stronger for
the second overtone. The variations of the two amplitudes are
anticorrelated: maximum amplitude of one mode always coincides with
the minimum amplitude of the other mode.

The discovery of modulated 1O/2O Cepheids shows that the Blazhko
effect and the double-mode pulsations are not mutually exclusive. It
also imposes very strong constraints on any proposed theoretical
model of the Blazhko effect. All three currently most popular models
(\cite{MORP,NRRP,Stothers}) fail to account for the properties of
these stars \cite{MK07}. The pattern of the modulation observed in
1O/2O Cepheids suggests that some form of energy transfer between
the two modes must be involved.

\subsection{Triple-mode Cepheids}
\label{sec:3.3}

Triple-mode Cepheids are extremely rare objects. Only 8 are know so
far, all have been found in the Magellanic Clouds
\cite{MKM04,LMCTripl,SMCCep}. They come in two different flavours:
they either pulsate in the fundamental mode, the first and the
second overtone (F/1O/2O type; 3 in LMC + 1 in SMC) or they pulsate
in the first three radial overtones (1O/2O/3O type; 2 in LMC + 2 in
SMC). All triple-mode pulsators are strongly dominated by the first
overtone, amplitudes of the other two modes are at least 3 times
lower.

\subsection{Secondary modes in RRab stars}
\label{sec:3.4}

Thanks to high quality photometry obtained with {\it Kepler} and
COROT space telescopes, secondary periodicities with mmag amplitudes
have been detected in many fundamental mode RR~Lyrae variables. A
group of 9 objects clearly stands out (Table~\ref{tab1}). In these
stars, the ratio of the secondary and the primary periods fall in a
very narrow range centered on 0.59. This is exactly the expected
period ratio of the second overtone and the fundamental mode,
$P_2/P_0$ (\cite{KepBenko}; Smolec, priv. comm). The stars of
Table~\ref{tab1} form a new group of double-mode radial pulsators:
the F/2O RR~Lyrae-type variables.

\begin{table}
\caption{F/2O double-mode RR Lyrae-type stars}
\label{tab1}
\begin{tabular}{lcclclccl}
\hline\noalign{\smallskip}
~star                      & ~$P_0$ [day] & ~$P_2/P_0$ & ~~ref.            & ~~~~~~~~~~ & star           & ~$P_0$ [day] & ~$P_2/P_0$ & ~~ref.            \\
\noalign{\smallskip}\svhline\noalign{\smallskip}
~V1127 Aql                 & ~0.3560      & ~0.5821    & ~~\cite{V1127}    & & V354 Lyr                  & ~0.5617      & ~0.5862    & ~~\cite{KepBenko} \\
~MW Lyr                    & ~0.3977      & ~0.5884    & ~~\cite{MWLyr}    & & COROT\thinspace 105288363 & ~0.5674      & ~0.5906    & ~~\cite{Corot1}   \\
~COROT\thinspace 101128793 & ~0.4719      & ~0.5837    & ~~\cite{Corot2}   & & V350 Lyr                  & ~0.5942      & ~0.5925    & ~~\cite{KepBenko} \\
~V2178 Cyg                 & ~0.4868      & ~0.5854    & ~~\cite{KepBenko} & & KIC\thinspace 7021124     & ~0.6225      & ~0.5931    & ~~\cite{KepNemec} \\
~V445 Lyr                  & ~0.5129      & ~0.5852    & ~~\cite{KepBenko} & &                           &              &            &                   \\
\noalign{\smallskip}\hline\noalign{\smallskip}
\end{tabular}
\end{table}

\subsection{Multi-mode radial pulsations: theory}
\label{sec:3.5}

\subparagraph{Linear theory: modeling pulsation periods}

Each measured period of an identified mode yields a very accurate
constraint on stellar parameters. In case of multi-mode radial
pulsators we have two or three such constraints, which makes these
objects particularly useful for testing stellar models. In the past,
analysis of double-mode Cepheids motivated the revision of stellar
opacities (\cite{Simon}). The observed periods of double-mode
variables can be used to derive metallicities of individual objects
(e.g. \cite{M33Cep,Buchler08}), and with additional input from
either the observed colours or from the evolutionary tracks, the
masses and luminosities of the stars or distances to stellar systems
can also be determined (e.g. \cite{Dekany08,Kovacs00}). With the
triple-mode pulsators, we can constrain stellar parameters even
further \cite{TriplSei}.

\subparagraph{Nonlinear models}

Double-mode pulsations turned out to be very resilient to
hydrodynamical modeling. The first nonlinear models displaying
stable full-amplitude double-mode behaviour were computed only after
time dependent turbulent convection was included into in the codes
\cite{beatmod1,beatmod2}. The results of these calculation have
recently been questioned by Smolec \& Moskalik \cite{SM08}. They
have shown that double-mode solutions found by
\cite{beatmod1,beatmod2} resulted from unphysical neglect of
buoyancy effects in convectively stable layers of the models. At
this point, the problem of reproducing the full-amplitude
double-mode pulsations is far from being solved.

\section{Nonradial modes in Cepheids}
\label{sec:4}

Resolved low amplitude secondary frequencies have been detected in
9\% of LMC first overtone Cepheids \cite{MK07}. In most cases they
are found very close to the frequency of the primary (radial) mode,
with $|\Delta f| < 0.13$\thinspace c/d. Similar secondary
periodicities have also been found in two F/1O double-mode Cepheids.
Close proximity of two frequencies cannot be reproduced with the
radial modes. Therefore, the secondary frequencies in these stars
must correspond to {\it nonradial modes}. Discovery of such modes
poses a challenge to the pulsation theory, which predicts that no
photometrically detectable nonradial modes should be excited in
Cepheid variables \cite{Mulet}.

\section{Mysterious period ratio of $P/P_1 \sim 0.62$}
\label{sec:5}

Secondary modes with puzzling period ratios in the range of
$0.600-0.645$ have been detected in more than 150 Magellanic Cloud
Cepheids \cite{MK07,LMCCep,SMCCep}. When plotted on the Petersen
diagram, these variables follow two (LMC) or three (SMC) well
defined parallel sequences. Secondary modes with almost the same
period ratios have also been detected in 13 RR~Lyrae stars
\cite{GranMos}. In both classes of stars, such modes are found only
in the first overtone and in the F/1O pulsators. The observed period
ratios are incompatible with excitation of two radial modes, neither
in Cepheids \cite{DS09}, nor in RR~Lyrae stars (Smolec, private
comm.). Therefore, the secondary frequency must be attributed to a
nonradial mode.

{\it Kepler} photometry of 4 RRc stars with $P/P_1\sim 0.62$
revealed another intriguing feature. In all 4 objects, subharmonics
of secondary frequencies have been detected. This means that the
secondary oscillations display a {\it period doubling}
\cite{GranMos}.

\begin{acknowledgement}
I gratefully acknowledge financial support from the conference
organizers. Work on this review was also supported in part by the
National Science Foundation under Grant No. NSF PHY05-51164.
\end{acknowledgement}

\end{document}